\documentclass[conference]{IEEEtran}

\usepackage{ifpdf}
\usepackage{cite}

\ifpdf
\usepackage[pdftex]{graphicx}
\else
\usepackage[dvips]{graphicx}
\fi

\usepackage{url}
\usepackage{hyperref}
\usepackage{xspace}
\usepackage{defs}

\begin{document}
\title{Spreadsheets for Stream Partitions and Windows}

\author{\IEEEauthorblockN{Martin Hirzel,
Rodric Rabbah,
Philippe Suter,
Olivier Tardieu and
Mandana Vaziri}
\IEEEauthorblockA{IBM T.J. Watson Research Center\\
Yorktown Heights, NY 10598\\
\textrm{\{}hirzel,rabbah,psuter,tardieu,vaziri\textrm{\}}@us.ibm.com}
}

\IEEEspecialpapernotice{Position Paper}

\maketitle

\begin{abstract}
We discuss the suitability of spreadsheet processors as tools for programming
streaming systems. We argue that, while spreadsheets can function as powerful
models for stream operators, their fundamental boundedness limits their scope of
application. We propose two extensions to the spreadsheet model and argue their
utility in the context of programming streaming systems.

\end{abstract}

\section{Stream Processing with Spreadsheets}
\label{sec:intro}
Stream processing is an appealing approach to problems involving high volumes
of data. Such problems arise in many different domains: finance, health care,
consumer analytics, telecommunications, etc. Streaming applications are
typically structured as a network of independent processing operators that
consume and produce \emph{tuples} of data. This architecture enables large-scale
deployments, as operators can be distributed and replicated across
machines for scalability, maximizing data-parallelism and thus resulting in
systems that can handle millions of tuples per second. It also naturally
enforces a clear separation of concerns: because operators only need to know
the format of incoming and outgoing tuples, their implementations are largely
independent of each other and they can even be developed by separate teams. 

We argue that spreadsheets are particularly well suited for programming
operators in streaming applications. Indeed, spreadsheet and stream
programming systems share many important commonalities: 
\begin{itemize}
    \item \textbf{Two-dimensional view of data}: spreadsheets organize data in rows
and columns, and streams are most naturally modelled as a sequence (rows) of
tuples of attributes (columns).
    \item \textbf{Reactive model of computation}: a change of value in a spreadsheet
cell results in the immediate propagation of computations to all dependent
cells. Similarly, stream operators produce new tuples as a reaction to inputs.
    \item \textbf{Stateless computations}: streaming systems embrace statelessness as it
facilitates encapsulation and thus parallelization. Spreadsheets favor
statelessness to maintain a model of idempotent (re-)computations.
\end{itemize}
A final point is simply pragmatism: domain experts are not always programmers,
but they will typically be familiar with spreadsheet processors, as more than
500~million people worldwide are reported to use spreadsheets
\cite{Gulwani11AutomatingStringProcessingSpreadsheets}. Supporting stream
programming within a familiar tool thus removes the need for an additional
expert programmer and the associated translation issues.

With these considerations in mind, we developed \activesheets
\cite{VaziriETAL14StreamProcessingSpreadsheet}, a platform for developing
streaming operators in Microsoft Excel. \activesheets integrates with IBM
Infosphere Streams and the SPL programming language \cite{HirzelETAL13SPL},
allowing end-users to visualize live streaming data in spreadsheets, and to
export their computation models as SPL operators.

As an illustration of programming with \activesheets, we consider a simple
financial application to detect bargains in quoted stock prices, based on
trading data. We observe a given stock symbol, with a history of trades at
prices $P_i$ and volumes $V_i$. The volume-weighted average price, a standard
metric in finance, is defined as:
\[ \mbox{VWAP} = \frac{\sum_i P_i \cdot V_i}{\sum_i V_i} \]
A quoted price is deemed a bargain if it is lower than the VWAP.

\begin{figure}
  \centerline{\includegraphics[width=\columnwidth]{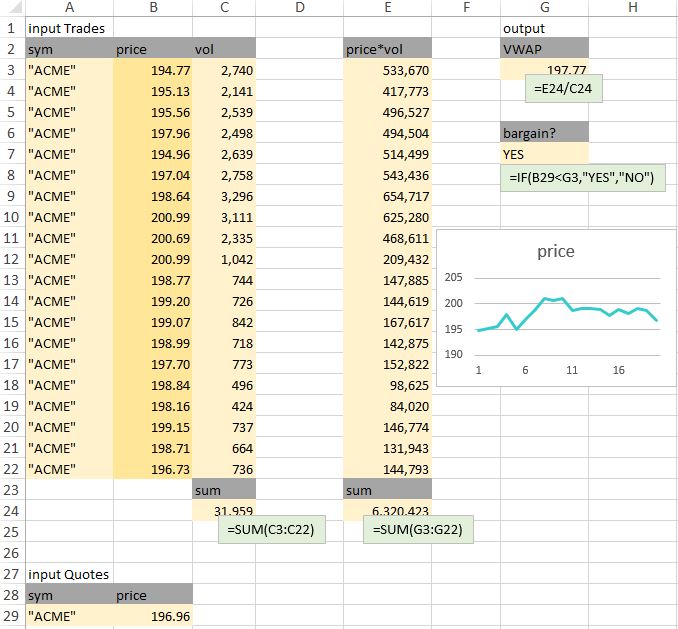}}
  \caption{A simple trading bargain calculator.}
  \label{fig:vwap}
  \vspace{-10pt}
\end{figure}
Figure~\ref{fig:vwap} shows how we can detect bargains in \activesheets. Cells
\xls{A3:C22} are mapped to the input stream of executed trades and, as tuples
come in, are updated to always reflect the last $20$ tuples (a scrolling
effect). This conforms with the natural view of streams as rows of tuples.
Cells \xls{A29:B29} contain the latest available quote, and are similarly
updated, potentially asynchronously from trades. All other cells and graphics
are standard Excel content, and update reactively to the incoming streams. The
\activesheets user marks the result of the computations for export, in this
case the cells \xls{G3} and \xls{G7}, thus determining the outputs of the
operator.

This programming experience is natural to users familiar with spreadsheets, and
leads to rapid development of malleable models. A fundamental
limitation of spreadsheets, however, is their bounded nature. In the following sections, we
highlight how this affects the programming of streaming operators and propose
potential extensions to alleviate these issues.

\section{Variable-size Windows}
We first consider a limitation arising from boundedness in the vertical
dimension. In the example described in the previous section, the computation of
the VWAP happens over a window of fixed size (the \emph{last $20$ values}). A
more realistic model would instead mandate aggregation over, e.g., the
\emph{last minute of trading data}. Such a window is problematic to represent
in a spreadsheet, as its size in terms of tuples cannot in general be computed
statically.

One approach (the current state of \activesheets) is to require application
developers to compute aggregations over variable-size windows outside of the
spreadsheet, in a separate set of operators, and to only stream the aggregated
data into the spreadsheet. While it can be straightforward for problems as
simple as VWAP, in general it takes away flexibility from the spreadsheet
developer and limits the expressiveness of the models that can be developed.
Another solution would be to impose a \emph{maximum} size on variable-size
windows. This would make it possible to allocate a sufficiently large number of
cells in the spreadsheet in advance, but would not properly support scenarios
where an unpredictably high number of tuples must be consumed in a short
timespan.

We argue that a better approach is to introduce a new function to
spreadsheets, \xls{WINDOW}, to allow users to store a \emph{list} of
values in a single cell. The spreadsheet processor can enforce that such
cells only flow into aggregating functions such as \xls{COUNT}, \xls{SUM},
\xls{AVERAGE}, etc., producing \xls{\#VALUE!} errors otherwise. The aggregating functions
are by design already capable of handling variable argument counts, and also have sensible
default values when applied to empty ranges. Their semantics
when applied to a window are thus predictable and natural.

Reactivity can be preserved by triggering recomputation only when a new value
enters the window. This may go counter to the notion that the window should
contain \emph{exactly} the values for, say, the last minute, but we believe it
matches the natural intuition that the value of a cell changes whenever its inputs do.
Finally, many such aggregation functions can be implemented efficiently using
incremental or partial recomputations, preserving the performance and
responsiveness associated with spreadsheets.

\section{Key-based Partitioning}
The second limitation we want to address is in handling unboundedness in data
variation. In our example in Figure~\ref{fig:vwap}, we detect bargains for the
ACME ticker symbol only, but what if we wanted to detect bargains for other
symbols, or for all incoming symbols?

Quotes and trades for a specific symbol can be easily extracted from the
streams of all quotes and trades using \xls{IF} or \xls{MATCH} expressions as
filters. One approach is therefore to replicate the computation $n$-ways,
selecting a particular symbol each time. While this approach is acceptable for a
small and known number of symbols, it does not scale and requires defining a
list of symbols to monitor a priori.

We argue that a better approach is to let users define and visualize the full
computation for \emph{one} defined symbol. To do so, they can filter the
incoming stream with a special \xls{SELECT} function, whose behavior is to
update the value of the cell only if its argument matches an expected value.
The rest of the computation is left as usual. Once the computation is fully
defined, the programmer replaces \xls{SELECT} by another new function,
\xls{PARTITION}. Visually, \xls{PARTITION} behaves identically. The semantics,
however, are that for each distinct value that flows into \xls{PARTITION}, a
conceptually independant instance of the spreadsheet performs the computation.
We can think of programming with \xls{PARTITION} as another instance of the
paradigm, familiar to spreadsheets users, of programming by example.

In the VWAP example, the \xls{PARTITION} function would be used on two streams:
quotes and trades. To preserve ease of use and enable efficient implementation,
we can require that all occurrences of the \xls{PARTITION} function in a
spreadsheet agree on the key, i.e., we can forbid partitioning the trades by
ticker symbol and the quotes by, say, geography.

\section{Conclusion}
Spreadsheets are a suitable medium for the development of operators in
streaming applications. Their familiar interface and computation model make it
possible for domain experts to participate in the development of complex
applications. However, natural operations on streaming data such as windowing
and partitioning are not directly expressible in spreadsheets due to their
bounded nature. These limitations restrict the autonomy of application
developers, thus partially defeating the purpose of the separation of concerns.
We propose extensions to the spreadsheet model that expand the scope of
expressible computations and that, we believe, adopt natural semantics.  We
plan to formalize our expanded computation model and to support the proposed
extensions in a future version of \activesheets.


\bibliographystyle{IEEEtran}
\bibliography{paper}

\end{document}